\documentclass[12pt]{article}
\pdfoutput=1
\usepackage{graphicx,amsfonts,amsthm,amsmath,amssymb,hyperref,color,yfonts,cite}
\newcommand{\Tr}{{\rm Tr}}
\newcommand{\bea}{\begin{eqnarray}\displaystyle}
\newcommand{\eea}{\end{eqnarray}}

\begin{document}
\makeatletter
\@addtoreset{equation}{section}
\makeatother
\renewcommand{\theequation}{\thesection.\arabic{equation}}
\vspace{1.8truecm}

{\LARGE{ \centerline{\bf Holography of a single free matrix}}}  

\vskip.5cm 

\thispagestyle{empty} 
\centerline{{\large\bf Robert de Mello Koch$^{a,b,d}$\footnote{{\tt robert@zjhu.edu.cn}}, Pratik Roy$^{b,d}$\footnote{{\tt roy.pratik92@gmail.com}} and
 Hendrik J.R. Van Zyl a$^{c,d}$\footnote{\tt hjrvanzyl@gmail.com}}}
\vspace{.8cm}
\centerline{{\it $^{a}$School of Science, Huzhou University, Huzhou 313000, China,}}
\vspace{.8cm}
\centerline{{\it $^{b}$School of Physics and Mandelstam Institute for Theoretical Physics,}}
\centerline{{\it University of the Witwatersrand, Wits, 2050, }}
\centerline{{\it South Africa }}
\vspace{.8cm}
\centerline{{\it $^{c}$The Laboratory for Quantum Gravity \& Strings,}}
\centerline{{\it Department of Mathematics \& Applied Mathematics,}}
\centerline{{\it University of Cape Town, Cape Town, South Africa}}
\vspace{.8cm}
\centerline{{\it $^{d}$ The National Institute for Theoretical and Computational Sciences,}} \centerline{{\it Private Bag X1, Matieland, South Africa}}

\vspace{1truecm}

\thispagestyle{empty}

\centerline{\bf ABSTRACT}

\vskip.2cm 
In this paper we consider the collective field theory description of a single free massless scalar matrix theory in 2+1 dimensions. The collective fields are given by $k$-local operators obtained by tracing a product of $k$-matrices. For $k=2$ and $k=3$ we argue that the collective field packages the fields associated to a single and two Regge trajectories respectively. We also determine the coordinate transformation between the coordinates of the collective field theory and the bulk AdS space time. This is used to verify that the bulk equations of motion holds in the collective field theory description.

\setcounter{page}{0}
\setcounter{tocdepth}{2}
\newpage
\tableofcontents
\setcounter{footnote}{0}
\linespread{1.1}
\parskip 4pt

{}~
{}~

\section{Introduction}

The free O(N) vector model, described by the action
\bea
S=\int d^3 x {1\over 2}\partial_\mu\phi^a\partial^\mu\phi^a
\eea
provides an instructive example of holography \cite{Maldacena:1997re,Gubser:1998bc,Witten:1998qj}. This conformal field theory is dual \cite{Klebanov:2002ja} to higher spin gravity \cite{Vasiliev:1990en}. An early suggestion \cite{Das:2003vw} for a constructive approach to developing holography suggested that the higher spin gravity description is obtained by expressing the dynamics of the conformal field theory in terms of invariant collective variables. A key insight motivating \cite{Das:2003vw} is the observation that although the loop expansion parameter of the original conformal field theory is $\hbar$, the loop expansion parameter of the holographic gravity theory is $1/N$. To change between these loop expansion parameters one should change to gauge invariant field variables, which is known as the collective field theory representation \cite{Jevicki:1979mb,Jevicki:1980zg}. The proposal of \cite{Das:2003vw} was explicitly realized in a light front quantization in \cite{deMelloKoch:2010wdf}. In this case the collective variables are given by equal $x^+$ bilocals
\bea
\sigma (x^+,x_1^-,x_1,x_2^-,x_2)&=&\sum_{a=1}^N\phi^a(x^+,x^-_1,x_1)\phi^a(x^+,x^-_2,x_2)
\eea
This bilocal field can be written as the sum of a large $N$ expectation value of the field, denoted $\sigma_0(x^+,x^-_1,x_1,x^-_2,x_2)$, plus a fluctuation as follows 
\bea
\sigma(x^+,x^-_1,x_1,x^-_2,x_2)&=&\sigma_0(x^+,x^-_1,x_1,x^-_2,x_2)
+{1\over\sqrt{N}}\eta(x^+,x^-_1,x_1,x^-_2,x_2)
\eea
It is the fluctuation $\eta(x^+,x^-_1,x_1,x^-_2,x_2)$ that is identified with the dynamical higher spin gravity fields. Indeed, after the change of coordinates
\bea
X&=& \frac{p_1^+ x_1+p_2^+ x_2}{p_1^++p_2^+}\qquad
Z\,\,=\,\,\frac{\sqrt{p_1^+ p_2^+} (x_1-x_2)}{p_1^++p_2^+}\qquad X^+=x^+\cr
P^+&=& p_1^++p_2^+\qquad\qquad
\theta\,\,=\,\,2 \tan ^{-1}\left(\sqrt{\frac{p_2^+}{p_1^+}}\right)
\label{coordmap}
\eea
the relation between the fields is given by
\bea
\Phi&=&\sum_{s=0}^\infty
\left(\cos (2s\theta) {A^{XX\cdots XX}(X^+,P^+,X,Z)\over Z}+\sin (2s\theta) {A^{XX\cdots XZ}(X^+,P^+,X,Z)\over Z}\right)\cr\cr
 &=&2\pi P^+\sin\theta\,\,  
\eta(X^+,P^+\cos^2{\theta\over 2},X+Z \tan{\theta\over 2},P^+\sin^2{\theta\over 2},X-Z\cot {\theta \over 2})
\eea
where $A^{XX\cdots XX}(X^+,P^+,X,Z)$ and $A^{XX\cdots XZ}(X^+,P^+,X,Z)$ are the two independent and physical components of the spin $2s$ gauge field in light cone gauge \cite{Metsaev:1999ui}. The above mapping is determined entirely by conformal symmetry and it provides a detailed matching between independent degrees of freedom in the conformal field theory and physical and independent degrees of freedom in the higher spin gravity \cite{deMelloKoch:2023ngh}. This gives a construction of the gravity fields that obey the correct gravity equations of motion \cite{deMelloKoch:2010wdf} with the correct boundary conditions \cite{Mintun:2014gua}, it explicitely realizes \cite{deMelloKoch:2021cni} entanglement wedge reconstruction \cite{Dong:2016eik} and it is manifestly \cite{deMelloKoch:2022sul} consistent with the principle of the holography of information \cite{Laddha:2020kvp}. The change of coordinates (\ref{coordmap}) can been recovered as a consequence of bulk locality \cite{deMelloKoch:2024juz}. Finally, the generalization of the map for $d$-dimensional conformal field theory was worked out in \cite{Jevicki:2011ss,Jevicki:2011aa} (see also \cite{Mintun:2014gua}).

Given this success, a natural extension is to consider matrix models. Specifically, we study the conformal field theory given by
\bea
S=\int d^3 x \,\,{1\over 2}\,\Tr \left(\partial_\mu\phi\partial^\mu\phi\right)
\eea
For the case of matrix models, the space of invariants is much larger. As in the vector model, we will have a bilocal invariant given by $\Tr(\phi(x_1)\phi(x_2))$. The primaries captured in the bilocal invariant are a scalar of dimension 1 and a family of conserved higher spin currents, one of each even integer spin. However, we now also have the possibility of trilocal $\Tr(\phi(x_1)\phi(x_2)\phi(x_3))$, quadlocal and in general, $k$-local field invariants. It is only the bilocal invariant that packages conserved currents. Recall that every single trace primary operator of the conformal field theory corresponds to a field in the AdS gravity, and further, primaries that are conserved currents correspond to gauge fields. Consequently the gauge fields of the dual gravity are packaged into the bilocal invariant, and all remaining invariants package massive fields of the gravity theory, which we will refer to as matter fields. The mapping for the gauge fields works exactly as in \cite{deMelloKoch:2010wdf}; the matter fields are a new feature of the free matrix conformal field theory. Our goal in this article is to understand how the holographic mapping works for these matter fields.

The first non-trivial question to answer is exactly what primary operators are packaged in each multi-local gauge invariant operator. We consider this question in Section \ref{bilocalholography}. Using conformal characters and respecting the cyclic symmetry of the trace, we are able to write Hilbert series for each gauge invariant operator. Expanding these series gives the dimension and spin of the relevant primary operators. The Hilbert series representation also suggests how these primaries can be constructed and in a few examples we explicitly carry this construction out.

Leaning heavily on the work of Metsaev \cite{Metsaev:2003cu,Metsaev:2005ws,Metsaev:2013kaa,Metsaev:2015rda,Metsaev:2022ndg}, we review the description of spinning fields in light front AdS in Section \ref{FieldsInAdS}. In Section \ref{HMapp} we discuss the holographic dual to the collective description. We argue that the fields packaged into the trilocal collective field correspond to two Regge trajectories, providing a compelling hint that there is underlying string theory dynamics. Using the requirement of bulk locality we then construct the mapping between the coordinates of the collective field theory and those of the AdS space time. This allows us to demonstrate that the bulk equations of motion found by Metsaev do indeed hold in the collective field theory. We collect conclusions and suggestions for further work in Section \ref{conclusions}.

\section{Single Trace Primary Operators}\label{bilocalholography}

In this section we study the single trace primary fields of the conformal field theory. Every single trace primary in the conformal field theory corresponds to a field on AdS$_4$ in the dual gravity theory. Determining the spectrum of single trace primaries in the conformal field theory determines the field content of the dual gravity theory. Primary operators, which correspond to irreducible representations of so(2,3), are labelled by a dimension $\Delta$ and a spin $j$. We label the irreducible representations by this pair of numbers $\Delta,j$.

Multilocal operators do not transform in an irreducible representation of so(2,3). Our goal is to take the representation relevant for a given multilocal operator and decompose it into irreducible representations, using the methods developed in \cite{deMelloKoch:2017caf,deMelloKoch:2017dgi,DeMelloKoch:2018hyq,deMelloKoch:2018klm}. This analysis uses character methods. See also \cite{Bae:2016hfy} for related relevant and useful ideas. The character that we need is given by
\bea
\chi_R (s,t)=\Tr_R (g)\qquad\qquad g=s^{2D}t^{J_3}
\eea
where $R$ is a representation of so(2,3), $\Tr_R$ is a trace over the states in this representation, $D$ is the dilatation operator and the so(3) subgroup of spatial rotations has basis $J_{\pm},J_3$. The basic character from which everything else is constructed is the character for the representation of the free scalar field. The free scalar field has dimension $\Delta={1\over 2}$. It transforms in a short representation of so(2,3) because it has null states corresponding to the free equation of motion and its descendants. The character of the free scalar field is
\bea
\chi_{({1\over 2},0)}(s,t)={s(1-s^4)\over (1-s^2 t)(1-s^2)(1-s^2/t)}
\eea
Conserved currents with spin $j$ have a dimension of $j+1$. This is again a short representation with null states corresponding to the conservation equation of the current. The character of a conserved current of spin $j$ is given by
\bea
\chi_{(j+1,j)}(s,t)&=& {s^{2j+2}(\chi_j(t)-s^2\chi_{j-1}(t))\over (1-s^2 t)(1-s^2)\left(1-s^2/t\right)}
\eea
where $\chi_j(t)$ is the character of the spin $j$ representation of so(3)
\bea
\chi_j(t)&=&\sum _{a=-j}^j t^a
\eea
Finally, the character of a representation of spin $j$ and dimension $\Delta>j+1$ is given by
\bea
\chi_{(\Delta,j)}(s,t)&=&{s^{2\Delta}\chi_j(t)\over (1-s^2 t)(1-s^2)\left(1-s^2/t\right)}
\eea
For a very readable account of these characters and how they are computed, the reader can consult \cite{Dolan:2005wy}. With these characters we can resolve the multilocal invariants into a sum over irreducible representations. Although the answers for the bilocal invariant is known, we will start with this case as a pedagogical device. After using character methods to extract the irreducible content we outline a convenient polynomial approach to the construction of the corresponding primaries \cite{deMelloKoch:2017caf,deMelloKoch:2017dgi,DeMelloKoch:2018hyq,deMelloKoch:2018klm}. Using this polynomial description, we explain how the known spinning currents are recovered. We then consider the trilocal invariant in detail. The trilocal invariant is sufficiently complicated that it illustrates the general case. In the context of the trilocal invariant we explain how Hilbert series provide a systematic approach to the problem of extracting the irreducible representations which appear and further, that it gives guidance on the construction of the corresponding polynomials. We also explain how the linear independence of the polynomials can be probed to ensure that we have indeed generated a complete set of primary operators. Using these tools we consider the quadlocal case in detail and also give some general remarks on the $k$-local case for general $k$.

\subsection{Bilocal collective fields}

The bilocal collective field is
\bea
\sigma_2 (x^+,x_1^-,x_1,x_2^-,x_2)=\Tr\left(\phi (x^+,x_1^-,x_1)\phi (x^+,x_2^-,x_2)\right)
\eea
Since the trace is cyclic, this operator is symmetric under swapping $1\leftrightarrow 2$, i.e. it is symmetric under the group ${\mathbb Z}_2=\{1,(12)\}$. The symmetric representation has characters $\chi_{{\mathbb Z}_2,{\rm sym}}(g)=1$ for $g\in {\mathbb Z}_2$. The projector onto the symmetric representation is
\bea
P_{\rm sym}={1\over |{\mathbb Z}_2|}\sum_{\sigma\in{\mathbb Z}_2}\chi_{{\mathbb Z}_2,{\rm sym}}(\sigma)\, \sigma = {1\over 2}\left(1+(12)\right)
\eea
where the permutations swap copies of the so(2,3) representation space of the scalar field. The character associated to the representation of the bilocal operator is now given by 
\bea
\chi_{{\rm Cyc}({1\over 2},0)^{\otimes 2}}&=&\Tr (P_{\rm sym}\, g\otimes g)\cr
&=&{1\over 2}\left(\Tr (g)^2+\Tr (g^2)\right)\cr
&=&{1\over 2}\left(\chi_{({1\over 2},0)}(s,t)^2+\chi_{({1\over 2},0)}(s^2,t^2)\right)
\eea
Using the characters quoted in the introduction to this section, some simple algebra proves that
\bea
{1\over 2}\left(\chi_{({1\over 2},0)}(s,t)^2+\chi_{({1\over 2},0)}(s^2,t^2)\right)&=&
\chi_{(1,0)}(s,t)+\sum_{j=1}^\infty\chi_{(2j+1,2j)}(s,t)
\eea
Thus, an analysis of the relevant characters has proved that
\bea
{\rm Cyc}\left(({1\over 2},0)^{\otimes 2}\right)=
(1,0)\oplus \bigoplus_{s=1}^\infty (2s+1,2s)\label{FFThm}
\eea
This result is the celebrated Flato–Fronsdal theorem\cite{Flato:1978qz,Basile:2014wua}. The right hand side above is a complete list of the primary operators that are collected into the bilocal collective field. It is this character analysis that we repeat below to learn about the primary operators packaged in the trilocal, quadlocal and $k$-local collective fields.

We would like an explicit construction of the primary fields we have just listed. One way to proceed is to use the following realization of the $({1\over 2},0)$ representation of so(2,3) \cite{deMelloKoch:2017caf,deMelloKoch:2017dgi}
\bea  
K_{\mu}&=&{\partial\over\partial x^{\mu}}\cr\cr 
P_{\mu}&=&\left(x^2\partial_\mu-2x_\mu x\cdot\partial -x_\mu\right)\cr\cr 
D&=&\left(x\cdot\partial+{1\over 2}\right)\cr\cr 
M_{\mu\nu}&=&x_{\mu } \partial_{ \nu} - x_{\nu} \partial_{\mu} \label{pgens}
\eea
To discuss an operator constructed from $n$ fields, we need the tensor product of $n$ copies of the $({1\over 2},0)$ representation. For the bilocal collective field we have $n=2$, but we will keep $n$ general for now. To obtain the tensor product of $n$ copies of the $({1\over 2},0)$ representation, introduce $n$ sets of coordinates $x^\mu+i$, $i=1,\cdots,n$ and use the operators
\bea  
K_{\mu}&=&\sum_{i=1}^n {\partial\over\partial x_i^{\mu}}\cr\cr 
P_{\mu}&=&\sum_{i=1}^n\left(x_i^2\partial_{i,\mu}-2x_{i,\mu} x_i\cdot\partial_i -x_{i,\mu}\right)\cr\cr 
D&=&\sum_{i=1}^n\left(x_i\cdot\partial_i+{1\over 2}\right)\cr\cr 
M_{\mu\nu}&=&\sum_{i=1}^n x_{i,\mu} \partial_{i,\nu} - x_{i,\nu} \partial_{i,\mu} \label{pgens}
\eea
With this representation, primary operators correspond to polynomials $\Psi$ in the $x_i^\mu$ that obey three conditions
\begin{eqnarray}
K_{\mu}\Psi &=&\sum_{i=1}^n\frac{\partial}{\partial x_i^{\mu}}\Psi =0\cr\cr
{\cal L}_{i}\Psi&=&\sum_{\mu}\frac{\partial}{\partial x_i^{\mu}}\frac{\partial}{\partial x_i^{\mu}}\Psi=0\cr\cr
\Psi (x^{\mu}_i)&=&\Psi (x^{\mu}_{\sigma (i)}) \qquad \sigma\in {\cal G}
\label{PrimaryConditions}
\end{eqnarray}
i.e. primary operators correspond to translation invariant, harmonic and symmetric polynomials. The choice of the group ${\cal G}={\mathbb Z}_n$ reflects the cyclic symmetry of the trace. If we had considered a product of scalar fields, for example, we would have to implement bosonic symmetry, so we would choose ${\cal G}=S_n$. 

For the bilocal operators we have two sets of coordinates $x^\mu_i$ with $i=1,2$. To build translation invariant polynomials, use the variables
\bea
y^\mu\equiv x^\mu_1-x^\mu_2
\eea
This ensures that we obey the first condition in (\ref{PrimaryConditions}). The third condition in (\ref{PrimaryConditions}) forces our polynomial to be even under swapping $x^\mu_1\leftrightarrow x^\mu_2$. Thus, our polynomials are of even degree in $y^\mu$. This ensures that we obey the third condition in (\ref{PrimaryConditions}) and it is the reason why we get only even spins in (\ref{FFThm}) and not all possible spins. To obey the second condition in (\ref{PrimaryConditions}), we need to consider harmonic polynomials, in which case the degree of the polynomial is the spin of the primary operator. If we use a polarization tensor $\epsilon^\mu$ that is traceless
\bea
\eta_{\mu\nu}\epsilon^\mu\epsilon^\nu =(\epsilon^0)^2-(\epsilon^1)^2-(\epsilon^2)^2=0
\eea
we find that the complete set of polynomials are given by
\bea
(\epsilon\cdot y)^{2s}=(\epsilon\cdot x_1-\epsilon\cdot x_2)^{2s}
\eea
Given the complete set of polynomials, we can construct the corresponding primary operators. To construct the primary operator from the polynomial, we use the following translation between polynomials and operators \cite{deMelloKoch:2017caf,deMelloKoch:2017dgi}
\bea
(\epsilon\cdot x)^k\to {1\over (2k-1)!!}(\epsilon\cdot \partial)^k\phi  \label{polytoprimary}
\eea
To derive this rule, compute the right hand side by applying the momentum operator given in (\ref{pgens}) a total of $k$ times, which gives
\bea
(\epsilon\cdot P)^k \, 1= (2k-1)!! (\epsilon\cdot x)^k
\eea 
We now find
\bea
(\epsilon\cdot y)^{2s}&=&(\epsilon\cdot x_1-\epsilon\cdot x_2)^{2s}\cr\cr
&=&\sum_{k=0}^{2s}(-1)^k {(2s)!\over (2s-k)! k!}(\epsilon\cdot x_1)^k(\epsilon\cdot x_2)^{2s-k}\cr\cr
&\to&\sum_{k=0}^{2s}(-1)^k {(2s)!\over 2^{2s} (2s-k)! k!(2k-1)!!(4s-2k-1)!!}(\epsilon\cdot \partial)^k\phi (\epsilon\cdot\partial)^{2s-k}\phi\cr\cr
&&\label{recovercurrent}
\eea
This can be compared to the higher spin currents evaluated for example in Section 2.1 of \cite{Giombi:2016hkj}. The higher spin currents are
\begin{eqnarray}
J_s(x^+,x^-,x,\alpha)
&=&J_{\mu_1\mu_2\cdots\mu_s}(x^+,x^-,x)\alpha^{\mu_1}\alpha^{\mu_2}\cdots \alpha^{\mu_s}\cr\cr
&=&\sum_{k=0}^{s}
\frac{(-1)^k\, :(\alpha\cdot\partial)^{s-k}\phi^a(x^+,x^-,x)\;(\alpha\cdot\partial)^{k}\phi^a(x^+,x^-,x) :}
{k!(s-k)!\Gamma(k+{1\over 2})\Gamma(s-k+{1\over 2})}
\cr
&&\label{scurrent}
\end{eqnarray}
To see the equality (up to an unimportant overall constant) use the identity
\bea
\Gamma\left(n+{1\over 2}\right)={\sqrt{\pi}(2n-1)!!\over 2^n}
\eea
This provides another proof of the tensor product formula (\ref{FFThm}) and it constructs the relevant primary operators.

\subsection{Trilocal collective fields}

The list of primary operators packaged by the trilocal collective field as well as the construction of these primaries are new results. The trilocal collective field is given by
\bea
\sigma_3 (x^+,x_1^-,x_1,x_2^-,x_2,x_3^-,x_3)=\Tr\left(\phi (x^+,x_1^-,x_1)\phi (x^+,x_2^-,x_2)\phi (x^+,x_3^-,x_3)\right)
\eea
Since the trace is cyclic, this operator is symmetric under cyclic permutations of $1,2,3$, i.e. it is symmetric under the group ${\mathbb Z}_3=\{1,(123),(132)\}$. Again, in the symmetric representation we have the characters $\chi_{{\mathbb Z}_3,{\rm sym}}(g)=1$ for all $g\in {\mathbb Z}_3$, which leads to the projector
\bea
P_{\rm sym}={1\over |{\mathbb Z}_3|}\sum_{\sigma\in{\mathbb Z}_3}\chi_{{\mathbb Z}_3,{\rm sym}}(\sigma)\, \sigma = {1\over 3}\left(1+(123)+(132)\right)
\eea
Consequently the character associated to the trilocal collective field is 
\bea
\chi_{{\rm Cyc}({1\over 2},0)^{\otimes 3}}(s,t)&=&{1\over 3}\left(\chi_{({1\over 2},0)}(s,t)^3+2\chi_{({1\over 2},0)}(s^3,t^3)\right)
\eea
A character analysis paralleling that of the bilocal collective field proves that
\bea
{\rm Cyc}\left(({1\over 2},0)^{\otimes 3}\right)&=& ({3\over 2},0)\oplus \bigoplus_{n=0}^\infty \left( (n+2)\left( {9\over 2} + 3n, 3n+3   \right) \oplus n\left( {9\over 2} + 3n, 3n+2   \right)\oplus   \right.   \cr
& & \oplus (n+1)\left( {9\over 2} + 3n-1, 3n+2   \right) \oplus (n+1)\left( {9\over 2} + 3n-1, 3n+1   \right) \oplus     \cr
& & \left.   \oplus (n+1)\left( {9\over 2} + 3n+1, 3n+4   \right) \oplus (n+1)\left( {9\over 2} + 3n+1, 3n+3   \right)    \right)   \label{coolresult}\cr\cr
&&
\eea
We will again use the polynomial language, so that we can construct the primary operators listed above. The result (\ref{coolresult}) implies that there are six families of polynomials with the following properties\footnote{If the polarizations $\epsilon^\mu$ are stripped off the polynomial to expose the indices of the $x_i^\mu$ coordinates, we are left with a multi-index tensor. The spin of the polynomial is the spin of this multi-index tensor. Thus, for example, the spin of a symmetric traceless tensor is simply given by the number of indices.}:
\begin{itemize}
\item[1.] $n+2$ independent polynomials of degree $3(n+1)$ and spin $3(n+1)$.
\item[2.] $n$ independent polynomials of degree $3(n+1)$ and spin $3(n+1)-1$.
\item[3.] $n+1$ independent polynomials of degree $3(n+1)-1$ and spin $3(n+1)-1$.
\item[4.] $n+1$ independent polynomials of degree $3(n+1)-1$ and spin $3(n+1)-2$.
\item[5.] $n+1$ independent polynomials of degree $3(n+1)+1$ and spin $3(n+1)+1$.
\item[6.] $n+1$ independent polynomials of degree $3(n+1)+1$ and spin $3(n+1)$.
\end{itemize}
Since we have 3 fields, we have 3 sets of coordinates $x_n^\mu$. There are two independent types of translation invariant combinations
\bea
y_1^\mu &=& x_1^\mu-x_2^\mu\qquad {\rm and\,\,\,permutations}\cr\cr
y_2^\mu &=& x_1^\mu+x_2^\mu-2x_3^\mu\qquad {\rm and\,\,\,permutations}
\eea
Employing the $y_i^\mu$ variables ensures that we immediately solve the first of (\ref{PrimaryConditions}). As for the bilocal, a null polarization tensor $\epsilon^\mu$ is used to satisfy the second of (\ref{PrimaryConditions}).  The polynomials are thus constructed using $\epsilon \cdot y_1$ and $\epsilon \cdot y_2$ as building blocks.  Judiciously chosen combinations satisfy the third of (\ref{PrimaryConditions}). There are two building blocks of degree three 
\begin{eqnarray}
P_1 & = & (\epsilon\cdot x_1 - \epsilon \cdot x_2)(\epsilon\cdot x_2 - \epsilon \cdot x_3)(\epsilon\cdot x_3 - \epsilon \cdot x_1)    \nonumber \\
P_2 & = & (\epsilon \cdot x_1 + \epsilon \cdot x_2 - 2 \epsilon \cdot x_3)(\epsilon \cdot x_2 + \epsilon \cdot x_3 - 2 \epsilon \cdot x_1) (\epsilon \cdot x_3 + \epsilon \cdot x_1 - 2 \epsilon \cdot x_2)    \nonumber
\end{eqnarray} 
as well as a building block of degree two  
\begin{equation}
Q = (\epsilon \cdot x_1 - \epsilon x_2)^2 + (\epsilon \cdot x_2 - \epsilon x_3)^2 + (\epsilon \cdot x_3 - \epsilon x_1)^2
\end{equation}
As a consequence of the identity
\begin{equation}
Q^3 = 54 (P_1)^2 + 2 (P_2)^2
\end{equation}
the requirement that our polynomials are linearly independent requires that only $Q^{n}$ for $n=0,1,2$ are employed. We label our polynomials as $f_{\gamma,j}$ where $\gamma$ tells us which of the six families our polynomial belongs to and $j$ indexes members in a family. Using these ingredients we can write the combinations
\begin{eqnarray}
f_{1,j}&=&(P_1)^{n+1-j}(P_2)^{j} \qquad j\,\,=\,\,0,1,\cdots n+1 \cr\cr
f_{3,j}&=&(P_1)^{n-j}(P_2)^{j} Q \qquad j\,\,=\,\,0,1,\cdots n\,\,\cr\cr    
f_{5,j}&=&(P_1)^{n-j}(P_2)^{j} Q^2 \qquad j\,\,=\,\,0,1,\cdots n \label{thefs}
\end{eqnarray}
To construct the remaining three families of polynomials, we need to introduce one more building block, of degree 2 and spin 1, given by
\bea
v^{\rho}&=&\epsilon^{\mu\nu\rho}\left(x_{1,\mu}x_{2,\nu}+x_{2\mu}x_{3\nu}+x_{3\mu}x_{1\nu}\right)\label{vdefn}
\eea
The remaining three families of polynomials are given by
\begin{eqnarray}
f_{2,j}&=&(P_1)^{n-j}(P_2)^{j}Q^2 (\epsilon\cdot v) \qquad j\,\,=\,\,0, 1, \cdots n\cr\cr f_{4,j}&=&(P_1)^{n-j}(P_2)^{j}(\epsilon\cdot v)\qquad j\,\,=\,\,0,1,\cdots n\cr\cr
f_{6,j}&=&(P_1)^{n-j} (P_2)^{j}Q(\epsilon\cdot v)\qquad j\,\,=\,\,0, 1,\cdots n \label{thegs}
\end{eqnarray}

Finally, it is worth making a comment about the constraints (\ref{PrimaryConditions}) that must be obeyed by polynomials corresponding to primary operators in the conformal field theory. If the building blocks each obey the first and third conditions, then any product of them will too. For the second condition it is not enough to check that each building block is harmonic since a product of harmonic polynomials is not necessarily harmonic. As a concrete example of this, although $\epsilon\cdot v$ is harmonic, $(\epsilon\cdot v)^n$ for $n>1$ is not. This is why no higher powers of $\epsilon\cdot v$ appear in our construction.

\subsubsection{Hilbert Series Analysis}

From the above analysis it is already clear that there is a considerable increase in the complexity of the complete set of polynomials corresponding to primary operators when moving from the bilocal to the trilocal case. For this reason it is useful to employ some of the powerful systematic tools developed for the description of rings of polynomials. The basic theory we will employ is provided by the Hilbert series. The Hilbert series can be computed algorithmically from the character of the tensor product. Its structure gives information about how many families of independent polynomials are needed, and the form of the generators of these families. Useful discussions directed towards physicists, that are pertinent to our analysis are \cite{deMelloKoch:2017dgi,deMelloKoch:2018klm,deMelloKoch:2022dpj}.

A generic so(2,3) character, for the representation built on a primary operator with dimension $\Delta$ and spin $j$, takes the form
\bea
\chi_{(\Delta,j)}(s,t)&=&{s^{2\Delta}\chi_j(t)\over P(s,t)}\label{generichar}
\eea
Here by ``generic'' we simply mean a representation that is not short. The short representations (given by $({1\over 2},0)$ and $(j+1,j)$ in our field theory) have null states which modify the numerator of (\ref{generichar}). Notice that the numerator of (\ref{generichar}) captures the contribution to the character of the primary itself, while the denominator encodes the contribution from the descendants. These descendants are generated by acting with the three momenta on the primary and there is one factor in the denominator for each momentum component.
Introduce the notation
\bea
P(s,t)=(1-s^2 t)(1-s^2)(1-s^2/t)
\eea
for the denominator and write
\bea
\chi_{{\rm Cyc}({1\over 2},0)^{\otimes k}}(s,t)&=&\sum_{\Delta,j \in S}
{s^{2\Delta}\chi_j(t)\over P(s,t)}
\eea
where $S$ is a list of pairs of dimensions and spins $(\Delta,j)$, one for each distinct primary operator that is encoded into the $k$-local operator. It is a simple manipulation to verify that\footnote{For this verification it is useful to keep the explicit form of the so(3) character
\bea
\chi_{j}(t)&=&t^j+t^{j-1}+\cdots+t^{-j+1}+t^{-j}
\eea
in mind.}
\bea
\chi_{{\rm Cyc}({1\over 2},0)^{\otimes k}}(s,t)\,P(s,t)\,(1-t^{-1})&=&\sum_{\Delta,j \in S}
s^{2\Delta}\,(t^j-t^{-j-1})
\eea
Keeping only the positive powers we obtain the Hilbert series
\bea
HS(s,t)&=&\left(\chi_{{\rm Cyc}({1\over 2},0)^{\otimes k}}(s,t)\,P(s,t)\,(1-t^{-1})\right)_{\ge}\cr\cr
&=&{1\over 2\pi i}\oint dz\,\, \chi_{{\rm Cyc}({1\over 2},0)^{\otimes k}}(s,z)\,P(s,z)\,(1-z^{-1})
{1\over z-t}\cr\cr
&=&\sum_{\Delta,j \in S} s^{2\Delta} t^j\label{forHS}
\eea
This last step of keeping only positive powers has been accomplished with the following identity from complex analysis \cite{Newton:2008au}
\bea
{1\over 2\pi i}\oint_{\cal C}dz {z^n\over z-a}&=& a^n\qquad\,\, {\rm if}\quad n\ge 0\cr\cr
&=&0\qquad\quad {\rm if}\quad n< 0
\eea
It is the middle equality in (\ref{forHS}) that is used to compute $HS(s,t)$ given the character $\chi_{{\rm Cyc}({1\over 2},0)^{\otimes k}}(s,t)$, for $k>2$.

The general structure of the Hilbert series is
\bea
HS(s,t)&=&{p(s,t)\over \prod_i (1-s^{\Delta_i}t^{s_i})}
\eea
Each factor in the denominator above corresponds to a generator of our polynomial ring\footnote{The fact that the primary operators of the conformal field theory of a free scalar field correspond to a polynomial ring was proved in \cite{deMelloKoch:2018klm}.}. The number $\Delta_i$ gives two times the degree of the generator and the number $s_i$ gives the spin of the generator. The numerator encodes relations between the generators (since products of generators may not be linearly independent) as well as the number of different polynomial families. As a simple example, the Hilbert series of the bilocal primaries is given by
\bea
HS(s,t)&=&\sum_{\Delta,j \in S} s^{2\Delta} t^j\,\,=\,\,\sum_{n=0}^\infty s^{2+4n}t^{2n}\,\,=\,\, {s^2\over 1-s^4 t^2}
\eea
The $s^2$ in the numerator is for $\Tr(\phi(0)\phi(0))$ which is a dimension 1 and spin 0 primary. This simply translates into 1 in the polynomial language since there are no derivatives of the fields. The denominator tells us that we have a single generator $((x_1-x_2)\cdot\epsilon)^2$ of dimension 2 and spin 2. As we have seen in (\ref{recovercurrent}), this correctly reproduces the spinning currents.

Let us now return to the trilocal collective field. Using the methods already outlined, we obtain the following Hilbert series\footnote{There maybe some manipulation required to get the Hilbert series into a useful form. This Hilbert series we are considering, for example, could also be written as
\bea
HS(s,t)=\frac{s^3 \left(1-s^2t+s^4(t+t^2)-s^6 t^2+s^8 t^3\right)}{\left(1-s^2 t\right) \left(1-s^6 t^3\right)}
\eea
This does not look correct because there is no generator of degree 1 and spin 1. For this reason we have multiplied by $1=(1+s^2 t)/(1+s^2 t)$ to obtain (\ref{nicer}).}
\bea
HS(s,t)=\frac{s^3 \left(1+s^4 t+s^6 t^3+s^{10} t^4\right)}{\left(1-s^4 t^2\right) \left(1-s^6 t^3\right)}\label{nicer}
\eea
This Hilbert series implies that the complete set of primaries are generated using a pair of generators. One of them is degree 2 and spin 2, the other is degree 3 and spin 3. The obvious candidates are
\bea
g_1&=&(\epsilon\cdot (x_1-x_2))^2+(\epsilon\cdot (x_2-x_3))^2+(\epsilon\cdot (x_3-x_1))^2\cr\cr
g_2&=&\epsilon\cdot (x_1+x_2-2x_3)\epsilon\cdot (x_2+x_3-2x_1)
\epsilon\cdot (x_3+x_1-2x_2)\label{HSgenerators}
\eea
The numerator tells us that there are four families of polynomials. The $s^3\times 1$ term corresponds to polynomials of the form
\bea
g_1^n\, g_2^m\,\qquad n,m=0,1,2,\cdots
\label{gg}
\eea
The $s^3\times s^4 t$ term is given by polynomials of the form
\bea
g_1^n\, g_2^m\, \epsilon\cdot v\qquad n,m=0,1,2,\cdots
\label{ggv}
\eea
where we recall the definition of $v^\mu$ from (\ref{vdefn}) above. The $s^3\times s^6 t^3$ term is given by polynomials of the form
\bea
g_1^n\, g_2^m\, Q_1\qquad n,m=0,1,2,\cdots
\label{ggq1}
\eea
where
\bea
Q_1&=&\epsilon\cdot(x_1-x_2)\,\epsilon\cdot(x_2-x_3)\,\epsilon\cdot(x_3-x_1)
\eea
Finally, the $s^3\times s^{10} t^4$ term corresponds polynomials of the form
\bea
g_1^n\, g_2^m\, Q_2\qquad n,m=0,1,2,\cdots
\label{ggq2}
\eea
where $Q_2$ is the following degree 5 and spin 4 polynomial
\bea
Q_2&=&\epsilon\cdot(x_1-x_2)\,\epsilon\cdot(x_2-x_3)\,\epsilon\cdot(x_3-x_1)\, \epsilon_\rho\epsilon^{\mu \nu \rho}  \left( x_{1,\mu} x_{2,\nu} + x_{2,\mu} x_{3,\nu} + x_{3,\mu} x_{1,\nu}    \right)
\eea

We have checked that the description provided in (\ref{thefs}),(\ref{thegs}) and the description provided in (\ref{gg}),(\ref{ggv}),(\ref{ggq1}),(\ref{ggq2}) are in complete agreement. The Hilbert series can be computed in an algorithmic fashion once the character describing the representations associated to the collective field has been evaluated. It then gives definite information about the number of generators and the relations obeyed by these generators. This additional input is helpful in constructing the complete set of polynomials corresponding to the primary operators associated to the collective field.

\subsubsection{Checking linear independence}

It is straightforward to check that polynomials corresponding to primary operators obey the conditions (\ref{PrimaryConditions}). To prove that the set of polynomials gives a complete description it is also necessary to argue that the polynomials are linearly independent. This requires introducing an inner product on the space of polynomials. Computing the matrix of inner product for the complete set of  polynomials and verifying that this matrix is full rank establishes their linear independence. In this subsection, following \cite{deMelloKoch:2018klm}, we will introduce a useful inner product for this analysis.
 
A natural inner product on polynomials in $x^\mu$ is defined by\footnote{Note that this inner product is not Lorentz covariant. It is however perfectly adequate for our purposes.}
\bea 
\langle x^{\mu_1}\cdots x^{\mu_k},x^{\nu_1}\cdots x^{\nu_k}\rangle &=& \sum_{\sigma\in S_k}\delta^{\mu_1,\nu_{\sigma (1)}}\delta^{\mu_2,\nu_{\sigma (2)}} 
\cdots\delta^{\mu_k,\nu_{\sigma (k)}}
\eea
Polynomials of different degree are orthogonal in this inner product. Note that the sum runs over all possible pairings between the $x^{\mu_i}$s and the $x^{\nu_i}$s, reminding one of Wick's theorem. Indeed, we can explicitly realize this inner product as a Gaussian integral
\bea
\langle f,g\rangle &=& \int_{-\infty}^\infty {dx^0\over\sqrt{\pi}} \int_{-\infty}^\infty {dx^1\over\sqrt{\pi}}\int_{-\infty}^\infty{dx^2\over\sqrt{\pi}} e^{-{1\over 2}\left( (x^0)^2+(x^1)^2+(x^2)^2\right)}\, :f: \, :g:
\eea
The normal ordered polynomials are obtained by subtracting out all self Wick contractions. A nice compact formula is
\bea
:f:=e^{-{1\over 2}\left((\partial_0)^2+(\partial_1)^2+(\partial_2)^2\right)}\, f
\eea
There is an obvious extension to polynomials in the multi-variable $x_n^\mu$ case as follows
\bea 
\langle x^{\mu_1}_{n_1}\cdots x^{\mu_k}_{n_k},x^{\nu_1}_{m_1}\cdots x^{\nu_k}_{m_k}\rangle &=& \sum_{\sigma\in S_k}\delta^{\mu_1,\nu_{\sigma (1)}}\delta_{n_1m_{\sigma(1)}}\cdots\delta^{\mu_k,\nu_{\sigma (k)}}\delta_{n_k m_{\sigma (k)}}\label{geninnerprod}
\eea
There is again an obvious realization of this inner product as a Gaussian integral.

Since polynomials of different degree are orthogonal and for the bilocal case we have a single polynomial at each degree, we know that in this case our polynomials are linearly independent and there is nothing further to check. For the case of the polynomials associated to primaries packaged in the trilocal collective field, there are multiple polynomials of the same degree and the check is non-trivial. For example, polynomials of type 1 and type 2 are of the same degree, equal to $3n+1$. Consider the complete set
\bea
P^{12}_A=\{f_{1, j},f_{2,j}\}
\eea
where $A$, which indexes distinct polynomials, runs over $2n+2$ values. We have verified that the matrix of inner products
\bea
M^{12}_{AB}=\langle P^{12 *}_AP^{12}_B\rangle
\eea
has rank $2n+2$, for low values of $n$, using mathematica. Similarly, using the sets
\bea
P^{34}_A=\{f_{3,j},f_{4,j}\}\qquad P^{56}_A=\{f_{5,j},f_{6,j}\}
\eea
where $A$ again runs over $2n+2$ values, we can compute the matrices
\bea
M^{34}_{AB}=\langle P^{34 *}_AP^{34}_B\rangle\qquad 
M^{56}_{AB}=\langle P^{56 *}_AP^{56}_B\rangle
\eea
We again find, with the help of mathematica and restricting to low values of $n$, that these are full rank. This establishes that the polynomials give a complete description.

It is equally easy to consider a parallel analysis for the polynomials constructed using the Hilbert series as a starting point. Again, for the classes of polynomials we have tested numerically, we confirm linear independence.

\subsubsection{Primaries}

The final step in the construction of the primaries entails translating the polynomials we have constructed into operators. We will employ the Hilbert series description of the polynomials. Towards this end, introduce the three variables
\bea
a=\epsilon\cdot x_1\qquad b=\epsilon\cdot x_2\qquad c=\epsilon\cdot x_3
\eea
The pair of generators (\ref{HSgenerators}) can then be written as
\bea
g_1&=&(a-b)^2+(b-c)^2+(c-a)^2\cr\cr
g_2&=&(a+b-2c) (b+c-2a)(c+a-2b)
\eea
A key formula that we need is the following power series expansion
\bea
g_1^n g_2^m &=& \sum_{p=0}^{2n+2m}\sum_{q=0}^{2n+3m-p}\alpha_{pq}\,\, a^p b^q c^{2n+2m-p-q}
\eea
where the coefficient $\alpha_{pq}$ can be obtained by applying the Multinomial Theorem
\bea
(x_1+x_2+\cdots +x_m)^n=\sum _{k_1+k_2+\cdots +k_m\,=n;\ k_1,k_2,\cdots ,k_m\geq 0}{n \choose k_1,k_2,\ldots ,k_m}\prod_{t=1}^{m}x_{t}^{k_{t}}
\eea
to $g_1^n$ and $g_2^m$, and then collecting terms. The primary operator then follows upon using (\ref{polytoprimary}). For example, the primary operator corresponding to the polynomial $g_1^n g_2^m$ is
\bea
\sum_{p=0}^{2n+2m}\sum_{q=0}^{2n+3m-p}\alpha_{pq}\,\, {(\epsilon\cdot\partial)^p \phi (\epsilon\cdot\partial)^q\phi (\epsilon\cdot\partial)^{2n+2m-p-q}\phi\over (2p-1)!! (2q-1)!!(2n+4m-2p-2q)!!}
\eea
which gives a new family of primary operators.

\subsection{Quadlocal}

The quadlocal collective field is
\bea
&&\sigma_4 (x^+,x_1^-,x_1,x_2^-,x_2,x_3^-,x_3,x_4^-,x_4)\cr\cr
&&\qquad
=\Tr\left(\phi (x^+,x_1^-,x_1)\phi (x^+,x_2^-,x_2)\phi (x^+,x_3^-,x_3)\phi (x^+,x_4^-,x_4)\right)
\eea
Since the trace is cyclic, this operator is symmetric under ${\mathbb Z}_4=\{1,(1234),(13)(24),(1432)\}$. The symmetric representation has characters $\chi_{{\mathbb Z}_4,{\rm sym}}(\sigma)=1$ for all $\sigma\in {\mathbb Z}_4$. Arguing as we have done above, the character for the representation of the quadlocal collective field is 
\bea
\chi_{{\rm Cyc}({1\over 2},0)^{\otimes 4}}&=&{1\over 4}\left(\chi_{({1\over 2},0)}(s,t)^4+\chi_{({1\over 2},0)}(s^2,t^2)^2+2\chi_{({1\over 2},0)}(s^4,t^4)\right)
\eea
A straightforward computation leads to the Hilbert series
\bea
HS(s,t)&=&\big(s^4-s^6 (1+t)+s^8 (1+2 t+t^2)+s^{12} (1+t+t^2-t^3)
-s^{10} (1+t-t^3)\cr\cr
&&-s^{14} (t-t^2-2 t^4)-s^{16} (2 t^2+t^4-t^5)+s^{18} (t^3-t^4-t^5-t^6)\cr\cr
&&-s^{20}(t^3-t^5-t^6)-s^{22} (x^4+2 x^5+x^6)+s^{24} (t^5+t^6)-s^{26} t^6\big)
\cr\cr
&&\times\frac{1}{\left(1-s^2\right) \left(1-s^8\right) \left(1-s^2 t\right) \left(1-s^4 t^2\right) \left(1-s^8 t^4\right)}
\eea
Although this Hilbert series is rather complicated, we can for example, truncate to the maximal twist terms which have the form $s^{2k}t^k$. After performing this truncation we obtain
\bea
HS^{\rm L.T.}(s,t)&=&\frac{s^4+2s^{10}x^3+s^{12}x^4}{(1-s^4 x^2)^2(1-s^8 x^4)}
\eea
The denominator indicates that we have three generators and the numerator indicates that there are four families of polynomials. Since we have 4 fields, we have 4 sets of coordinates $x_n^\mu$. There are three independent translation invariant structures
\bea
y_1^\mu = x_1^\mu-x_2^\mu\qquad{\rm and\,\,permutations}\cr\cr y_2^\mu=x_1^\mu+x_2^\mu-2x_3^\mu\qquad{\rm and\,\,permutations}\cr\cr  y_3^\mu=x_1^\mu+x_2^\mu+x_3^\mu-3x_4^\mu\qquad{\rm and\,\,permutations}
\eea
From these we can build the following generators
\bea
g_1&=&(\epsilon\cdot (x_1-x_2))^2+(\epsilon\cdot (x_2-x_3))^2+(\epsilon\cdot (x_3-x_4))^2+(\epsilon\cdot (x_4-x_1))^2\cr\cr
g_2&=&(\epsilon\cdot (x_1+x_2-2x_3))^2+(\epsilon\cdot (x_2+x_3-2x_4))^2+(\epsilon\cdot (x_3+x_4-2x_1))^2\cr\cr
&&\quad+(\epsilon\cdot (x_4+x_1-2x_2))^2\cr\cr
g_3&=&\epsilon\cdot(x_1+x_2+x_3-3x_4)\,\epsilon\cdot(x_2+x_3+x_4-3x_1)\,\epsilon\cdot(x_3+x_4+x_1-3x_2)\cr\cr
&&\qquad\times \epsilon\cdot(x_4+x_1+x_2-3x_3)
\eea
The four families of polynomials corresponding to the maximal twist primaries are then given by
\bea
g_1^{n_1}g_2^{n_2}g_3^{n_3}Q_k\label{quadlocalpolys}
\eea
where
\bea
Q_1&=&1\cr\cr
Q_2&=&(\epsilon\cdot (x_1-x_2))^3+(\epsilon\cdot (x_2-x_3))^3+(\epsilon\cdot (x_3-x_4))^3+(\epsilon\cdot (x_4-x_1))^3\cr\cr
Q_3&=&(\epsilon\cdot (x_1+x_2-2x_3))^3+(\epsilon\cdot (x_2+x_3-2x_4))^3+(\epsilon\cdot (x_3+x_4-2x_1))^3\cr\cr
&&\quad+(\epsilon\cdot (x_4+x_1-2x_2))^3\cr\cr
Q_4&=&(\epsilon\cdot (x_1+x_2-2x_3))^4+(\epsilon\cdot (x_2+x_3-2x_4))^4+(\epsilon\cdot (x_3+x_4-2x_1))^4\cr\cr
&&\quad+(\epsilon\cdot (x_4+x_1-2x_2))^4
\eea
As a non-trivial test of this construction we have verified that the set of polynomials defined in (\ref{quadlocalpolys}) are linearly independent in the inner product (\ref{geninnerprod}) up to and including degree 10.

\subsection{$k$-local}

The $k$-local collective field is given by
\bea
\sigma_k (x^+,x_1^-,x_1,x_2^-,x_2,\cdots,x_k^-,x_k)=\Tr\left(\phi (x^+,x_1^-,x_1)\phi (x^+,x_2^-,x_2)\cdots \phi (x^+,x_k^-,x_k)\right)
\eea
Cyclicity of the trace implies that this operator is symmetric under cyclic permutations of $1,2,\cdots,k$, i.e. it is symmetric under the group ${\mathbb Z}_k$ generated by the $k$-cycle $(123\cdots k)$. The symmetric representation has characters $\chi_{{\mathbb Z}_k,{\rm sym}}(\sigma)=1$ so that the character associated to the representation of the $k$-local operator is 
\bea
\chi_{{\rm Cyc}({1\over 2},0)^{\otimes k}}&=&{1\over k}\sum_{\sigma\in {\mathbb Z}_k}\Tr(\sigma g^{\otimes k})
\eea
Restricting ourselves to prime $k$, we know that the group ${\mathbb Z}_k$ has a single element corresponding to the identity permutation and $k-1$ elements that are given by $k$-cycles. Consequently, we have the simple and explicit formula
\bea
\chi_{{\rm Cyc}({1\over 2},0)^{\otimes k}}&=&{1\over k}\left(\chi_{({1\over 2},0)}(s,t)^k +(k-1)\chi_{({1\over 2},0)}(s^k,t^k)\right)
\eea
In the previous subsection we exhibited a dramatic simplification of the Hilbert series upon focusing on the maximal twist primary operators. In this section we follow the same strategy and construct the Hilbert series for this maximal twist case. The leading twist contributions are the descendants of a single component of momentum. Consequently, we can simplify
\bea
\chi_{({1\over 2},0)}(s,t)&\to& \chi^{L.T.}_{({1\over 2},0)}(s,t)={s\over 1-s^2 t}
\eea
and
\bea
P(s,t)&\to& P^{L.T.}(s,t)\,\,=\,\,(1-s^2t)
\eea
The Hilbert series describing the maximal twist primary operators is then given by
\bea
HS^{L.T.}(s,t)&=&{1\over k}\left(\chi^{L.T.}_{({1\over 2},0)}(s,t)^k +(k-1)
\chi^{L.T.}_{({1\over 2},0)}(s^k,t^k)\right)P^{L.T.}(s,t)\cr\cr
&=&\frac{s^k \left(1+(k-1) \left(1-s^2 x\right)^k-s^{2 k} x^k\right)}{k \left(1-s^2 x\right)^{k-1} \left(1-s^{2 k} x^k\right)}
\eea
This formula is easily used to generate the spectrum of irreducible representations that appear for any prime $k$. To obtain some information about the ring of maximal twist primary operators, it is helpful to rewrite the Hilbert series as
\bea
HS^{L.T.}(s,t)&=&\frac{s^k p_k(s,x)}{k \left(1-s^4 x^2\right)^2 \left(1-s^8 x^4\right)^{k-4} \left(1-s^{2 k} x^k\right)}\label{MaxTwsitPrim}
\eea
where
\bea
p_k(s,x)=\left(s^2 x+1\right)^{k-2} \left(s^4 x^2+1\right)^{k-4} \left(\sum _{a=0}^{k-1} s^{2 a} x^a+(k-1) \left(1-s^2 x\right)^{k-1}\right)
\eea
An important property of the numerator $p_k(s,x)$, established through explicit computation with mathematica, is that it admits an expansion
\bea
p_k(s,x)=\sum_{q=0}^{4k-11} \alpha_q s^{2q}x^q
\eea
with positive coefficients $\alpha_q\ge 0$. Consequently the Hilbert series (\ref{MaxTwsitPrim}) implies that the ring of maximal twist primaries is a polynomial ring with $k-1$ generators. The fact that there are no negative signs in $p_k(s,x)$ implies that these generators do not satisfy any relations. For $k>5$ there is a total of $4k-12$ families of polynomials.  To understand why there are $k-1$ generators, note that the fact that we consider fields at $k$ distinct points, implies that we have $k$ distinct sets of coordinates $x_n^\mu$. There are $k-1$ independent translation invariant combinations that we can form
\bea
y_n^\mu = \sum_{i=1}^n x_i^\mu- n x_{n+1}^\mu\qquad n=1,2,\cdots,k-1
\eea
which matches the number of generators in the ring. The generators will be constructed as a function of these translation invariant combinations in much the same way that the construction of the primaries from the quadlocal collective field worked.

\section{Spinning Fields in AdS}\label{FieldsInAdS}

In light-cone gauge a complete gauge fixing of free massless higher spin gravity fields was achieved in \cite{Metsaev:1999ui}. This was extended to massive fields in \cite{Metsaev:2003cu,Metsaev:2005ws,Metsaev:2013kaa,Metsaev:2015rda,Metsaev:2022ndg}. Since we will make extensive use of these results in the next section, we need to review this background material.

Working in AdS$_4$, each field is associated with a positive energy unitary lowest weight representation of SO(2,3). These representations are labelled by a dimension ($\Delta$) and a spin ($s$). Fields for which $\Delta=s+1$ are in short representations and correspond to massless fields. The bulk fields dual to the bilocal collective field all have $\Delta=s+1$ and they correspond to a bulk scalar and even spin gauge fields. The bulk fields dual to the trilocal collective field all have $\Delta>s+1$ and these correspond to massive fields. It is this case of massive fields that is of interest to us below. 

The list of dimensions and spins of primary operators that we have obtained for both the trilocal and quadlocal collective fields can be translated into a spectrum of masses $m^2$ of fields in AdS$_4$. According to Metsaev (see for example formula (2.23) of \cite{Metsaev:2005ws}) a primary of spin $s$ and dimension $\Delta$ is dual to a massive field of mass squared $m^2$ with
\bea
m^2=\left(\Delta-{3\over 2}\right)^2-\left(s-{1\over 2}\right)^2
\eea
Recall from (\ref{coolresult}) that the trilocal collective field packages primaries with dimension and spin given by
\bea
\left( \Delta,s\right)&=&\left\{ \left(3n+{7\over 2},3n+2\right),\left(3n+{9\over 2},3n+3\right),\left(3n+{11\over 2},3n+4\right)\right\}
\eea
with $n=0,1,2,\cdots$. These dimensions and spins lead to a mass squared spectrum $m^2=k+0.75$ with $k=1,2,3,\cdots$, so that these primaries lie on the Regge trajectory
\bea
m^2&=&s-0.25\label{regge1}
\eea
providing clear evidence for an underlying string description. The scalar primary with $\Delta={3\over 2}$ has $m^2=-{1\over 4}$ so that it too lies on this Regge trajectory. The trilocal collective field also packages primaries with dimension and spin given by
\bea
\left( \Delta,s\right)&=&\left\{ \left(3n+{7\over 2},3n+1\right),\left(3n+{9\over 2},3n+2\right),\left(3n+{11\over 2},3n+3\right)\right\}
\eea
with $n=0,1,2,\cdots$. These primaries lie on a second Regge trajectory given by
\bea
m^2&=&3s+0.75\label{regge2}
\eea
Thus the complete set of primary operators packaged into the trilocal collective field fill out two Regge trajectories. 

Metsaev has also given the light cone wave equation that each field must obey
\bea
\left(2\partial^+\partial^-+\partial_X^2+\partial_Z^2-{A\over Z^2}\right)|\phi\rangle &=&0
\label{MEoM}
\eea
where $|\phi\rangle$ packages the spin degrees of freedom using oscillators $\alpha^I$, with $I$ ranging over $Z$ and the directions transverse to the light cone (for AdS$_4$ this is just $X$). In this way we are decomposing the so($d-1$)=so(3) totally symmetric and traceless representation into a sum over so($d-2$)=so(2) representations. Concretely we have\footnote{We employ the very useful oscillator representation introduced in \cite{Metsaev:1999ui}.}
\bea
|\phi\rangle&=&\sum_{s'=0}^s\oplus\alpha_{I_1}\cdots\alpha_{I_{s'}}\phi^{I_1\cdots I_{s'}}|0\rangle\label{defphi}
\eea
The oscillators $\alpha^I$ are creation operators and the corresponding annihilation operators are denoted $\bar{\alpha}^I$. They obey
\bea
[\alpha^I,\bar{\alpha}^J]&=&\delta^{IJ}\qquad\qquad \bar{\alpha}^I|0\rangle\,\,=\,\,0
\eea
The operator $A$, which acts only on the spin indices, is the AdS mass operator. It can be written in terms of $\alpha^I$ and $\bar{\alpha}^I$. A formula for the AdS mass operator is
\bea
A&=&2M^{ZX}M^{ZX}+2+\Delta (\Delta-3)+s(s+1)+2B^Z\cr\cr
&=&2B+2+\Delta (\Delta-3)+s(s+1)
\label{AdSMassOp}
\eea
In the next section we obtain formulas for the AdS bulk coordinates in terms of the coordinates of the collective description, and in the Appendix \ref{App1} we obtain the form of the  AdS mass operator in the collective description. With these results in hand we are able to directly confirm the equation of motion (\ref{MEoM}) holds in the collective field theory description.

\section{Holographic Mapping}\label{HMapp}

The requirement of bulk locality can be used to derive the coordinate transformation of bilocal holography, which relates the coordinates of the boundary conformal field theory to those of the bulk \cite{deMelloKoch:2024juz}. We will use this logic to determine the mapping relevant for the $k$-local collective field in a $d$-dimensional conformal field theory. We use light cone coordinates in the dual AdS$_{d+1}$ bulk. The coordinates of the AdS space time, in Poincar\'e patch are $(X^+,X^-,\vec{X},Z)$ with $\vec{X}$ a $d-2$ dimensional vector of coordinates transverse to the light cone. $Z$ is the usual holographic direction.

Consider a collective field dual to a bulk operator which is localized to the set of bulk points given by $X^+=0$, and definite values for $\vec{X}$ and $Z$. $X^-$ is not fixed so we are considering an operator localized to a light like line of points in the AdS$_{d+1}$ bulk. For this line of points, using the results of \cite{Metsaev:1999ui}, the momentum $P^+$ and special conformal transformation $K^+$ become
\bea
P^+&=&\partial^+\qquad\qquad K^+\,\,=\,\,-{1\over 2}(\vec{X}\cdot\vec{X}+Z^2)\partial^+
\eea
The combination $K^\mu+{1\over 2}(\vec{X}\cdot\vec{X}+Z^2)P^\mu$ therefore leaves this light like line of points fixed. This implies a non trivial differential equation that a bulk field localized on the light like line must obey
\bea
\big( K^++{1\over 2}(\vec{X}\cdot\vec{X}+Z^2)P^+\big)O_\Psi=0\label{exprsblkloc}
\eea
This equation is general, holding for fields of any spin, thanks to the form of $K^+$ given in (3.71) of \cite{Metsaev:1999ui}. A Fourier transform on $X^-$ in gravity replaces the differential operator $\partial^+$ with the variable $P^+$. Since we have localized only to the light like line parametrized by $X^-$, the Fourier transformed field obeys the same bulk locality condition. The results of \cite{deMelloKoch:2010wdf} imply that the coordinate transformation between the coordinates of the conformal field theory and the bulk AdS space time can be determined by matching the generators of conformal transformations of the bilocal conformal field theory with those of the higher spin gravity. This implies that we can insert the bilocal generators into the above equation. As usual we identify $X^+=x^+$. Inserting the bilocal expression for the generators into (\ref{exprsblkloc}), we obtain
\bea 
\Big(-{1\over 2}\sum_{i=1}^k \vec{x}_i\cdot\vec{x}_i p^+_i\,+\,{1\over 2}(\vec{X}\cdot\vec{X}+Z^2)\sum_{l=1}^k p_k^+\Big)O_\Psi&=&0\label{BlkEqn}
\eea
Notice that this is a polynomial multiplied by the field. Since the field does not vanish the polynomial does. Setting the polynomial multiplying $O_\Psi$ in (\ref{BlkEqn}) to zero relates the bulk AdS$_{d+1}$ coordinates $\vec{X}$ and $Z$ to the coordinates
$\vec{x}_i$ and $p_i^+$ of the conformal field theory. Recall that $\vec{x}_i$ are $d-2$ dimensional vectors of coordinates transverse to the light cone.

Matching the symmetry $X^-\to X^-+a$ in the bulk to $x_i^-\to x_i^-+a$ in the collective field theory we find that
\bea
P^+&=&\sum_{i=1}^k p_i^+
\eea
Denote the momentum conjugate to $\vec{X}$ by $\vec{P}$, and that conjugate to $\vec{x}_k$ by $\vec{p}_k$. An identical argument matching translations in the directions transverse to the light cone implies that
\bea
P^i=\sum_{j=1}^k p_j^i
\eea
Now consider an SO($d$-2) transformation $R^i{}_j$ which acts as
\bea
X^i\to R^i{}_j X^j
\eea
in the bulk, and as 
\bea
x_l^i\to R^i{}_j x_l^j
\eea
in the collective field theory. Thus, $\vec{X}$ and $\vec{x}_l$ are all in the $d-2$ dimensional vector representation of SO($d$-2). This suggests a linear relation between them
\bea
\vec{X}=\sum_{l=1}^k \alpha_l\vec{x}_l\qquad \sum_{l=1}^k\alpha_l =1
\eea
The condition that the coefficients $\alpha_l$ sum to 1 ensures that the translation $\vec{X}\to\vec{X}+\vec{a}$ in the bulk corresponds to the simultaneous translation $\vec{x}_l\to\vec{x}_l+\vec{a}$ in the collective field theory. The arguments we have considered so far allow the parameter $\alpha_l$ to be arbitrary functions of the momenta $p_j^+$. The bulk coordinate $Z$ is invariant under translations $X^i\to X^i+a^i$ and under the SO($d$-2) transformations $R^i{}_j$. This implies that $Z$ is a function only of $|\vec{x}_1-\vec{x}_2|$. Setting the polynomial multiplying $O_\Psi$ in (\ref{BlkEqn}) to zero now implies that
\bea
\vec{X}&=&{\sum_{i=1}^k p_i^+\vec{x}_i\over \sum_{j=1}^k p_j^+}
\qquad
Z\,\,=\,\,{\sqrt{\sum_{i=1}^kp_i^+ \vec{v}_i\cdot\vec{v}_i}\over \left(\sum_{j=1}^k p_j^+\right)^{3\over 2}}
\eea
where
\bea
\vec{v}_i&=&\sum_{j=1}^k p_j^+(\vec{x}_i-\vec{x}_j)
\eea
Notice that $Z=0$ requires that we set $\vec{v}_i=0$ for all $i$. Since it is easy to check that
\bea
\vec{x}_i-\vec{x}_j={\vec{v}_i-\vec{v}_j\over\sum_{l=1}^k p_l^+}
\eea
we see that $Z=0$ requires that all of the spatial points $\vec{x}_i$ in the $k$-local collective field become coincident. This observation implies that the arguments of \cite{deMelloKoch:2022sul} which show that bilocal holography realizes the holography of information \cite{Laddha:2020kvp,Chowdhury:2020hse,Raju:2020smc,Raju:2021lwh,SuvratYouTube}. For further background motivating the holography of information the reader can consult \cite{Marolf:2008mf,Jacobson:2012ubm,Papadodimas:2012aq,Banerjee:2016mhh,Raju:2018zpn,Raju:2019qjq,Jacobson:2019gnm}. We will be content with sketching the basic idea. Start with the observation that $k$-local collective fields, with the $k$ locations $\vec{x}_j$ well separated, correspond to fields localized deep in the bulk i.e. at a large value for $Z$. Using the operator product expansion (OPE) the $k$-local collective field can be reduced to a sum over conformal primaries and their descendants. These can all be constructed from the $k$-local collective field with arbitrarily small separations $|\vec{x}_j-\vec{x}_i|<\epsilon$. The $k$-local collective fields with arbitrarily small separations are all located in an arbitrarily small neighbourhood of the boundary. In this way all of the information available on any given Cauchy slice is available in an arbitrarily small neighbourhood of the boundary of the slice. To really construct a compelling argument, one would have to show that the relevant OPEs converge.

In what follows we will also need derivatives with respect to the AdS coordinates. These are given by
\bea
{\partial\over\partial X^i}&=&\sum_{j=1}^k {\partial\over\partial x^i_j}\qquad\qquad\qquad {\partial\over\partial Z}\,\,=\,\,{1\over Z}{\sum_{i=1}^k \vec{v}_i\cdot {\partial\over\partial \vec{x}_i}\over \sum_{j=1}^k p_j^+}\cr\cr
{\partial\over\partial P^+}&=&{\sum_{i=1}^k p_i^+{\partial\over\partial p_i^+}\over\sum_{j=1}^k p_j^+}\qquad\qquad
{\partial\over\partial X^+}={\partial\over\partial x^+}
\eea
Using the expressions for $M^{XZ}$ and $B^Z$ derived in Appendix \ref{App1}, we obtain the following translation of the bulk equation of motion into the collective description
\bea
\Big(C_2-\Delta (\Delta-3)-s(s+1)\Big) {\cal O}&=&0\label{CFTeom}
\eea
where $C_2$ is the quadratic Casimir of the conformal field theory ($M,N$ are summed over $-1,0,1,2,3$)
\bea
C_2&=&{1\over 2}L_{MN}L^{NM}\label{quadcas}
\eea
where we write $L_{MN}=-L_{NM}$ in terms of the conformal generators as
\bea
L_{\mu\nu}&=&J_{\mu\nu}\qquad L_{\mu\, -1}\,\,=\,\,{1\over 2}(P_\mu+K_\mu)\cr\cr
L_{-1\,3}&=&-D\qquad\quad L_{\mu 3}\,\,=\,\,{1\over 2}(P_\mu-K_\mu)
\eea
The indices $M$ and $N$ are raised and lowered with the metric $\eta={\rm diag}(-1,-1,1,1,1)$ and the conformal generators are listed in Appendix \ref{App2}.  Notice that the bulk equation of motion simply becomes an equation which selects the correct representation from the collective field. Therefore, it is obeyed in the collective field theory and it states precisely which primary operators are to be identified with which bulk fields.

\section{Discussion and Conclusions}\label{conclusions}

In this paper we have started to study the collective field theory of a free matrix in $2+1$ dimensions. The first step in this procedure is to formulate the dynamics in terms of (gauge invariant) collective fields, which are $k$-local fields for $k=1,2,3,\cdots$. For $k>1$ this collective field packages an infinite number of primary operators. We have explained how the methods of \cite{deMelloKoch:2017caf,deMelloKoch:2017dgi,DeMelloKoch:2018hyq,deMelloKoch:2018klm} allow us to work the list of spins and dimensions of the primary operators that appear. Further we have explained how these primary operators can be constructed.

Using the well developed light cone formulation of Metsaev \cite{Metsaev:2003cu,Metsaev:2005ws,Metsaev:2013kaa,Metsaev:2015rda,Metsaev:2022ndg} we have shown that the primaries packaged into the trilocal collective field lie on a pair of Regge trajectories, given in equations (\ref{regge1}) and (\ref{regge2}). Further, the primaries packaged into the bilocal collective field all lie on the Regge trajectory $m^2=0$. The fact that Regge trajectories appear is a strong hint of an underlying string theory description.

We have also described the coordinate transformation which related the coordinates of the collective field theory to the dual AdS spacetime. Formulas for the coordinates of the AdS$_{d+1}$ space time, needed to map $k$-local collective fields of CFT$_d$ have been obtained for any $d$. We have also verified that the bulk equation of motion has a natural interpretation in the collective field theory: it simply becomes the equation that isolates the relevant conformal primary, using the quadratic Casimir.

According to the coordinate transformation we have found, the value of the extra holographic coordinate $Z$ becomes larger as the separations in the $k$-local collective fields increase. The value $Z=0$ is only achieved when all $k$ fields are coincident. As we have argued above, this is a very strong hint that the collective description localizes information exactly as predicted by the principle of the holography of information \cite{Laddha:2020kvp,Chowdhury:2020hse,Raju:2020smc,Raju:2021lwh,SuvratYouTube} and hence exactly as expected in a theory of quantum gravity.

There are a number of interesting directions in which this work can be extended. A rather simple but instructive exercise would involve computing the spectrum of primaries for $k$-local collective fields with $k>3$. One could also consider the spectrum of primaries for $k$-local collective fields for more general values of $d$. A further instructive exercise is to demonstrate the matching of the equation of motion for $k>3$ and more general values of $d$.

Another nice direction would be to determine the complete coordinate transformation for the trilocal collective field. This would involve finding the transformation for three additional angles, which would be used to package the complete set of primaries into a single field. With this complete map, one could verify the GKPW boundary condition holds for the collective construction and thereby confirm that we have a complete bulk reconstruction.

\begin{center} 
{\bf Acknowledgements}
\end{center}
RdMK would like to thank Antal Jevicki for very helpful discussions. RdMK is supported by a start up research fund of Huzhou University, a Zhejiang Province talent award and by a Changjiang Scholar award. PR is also supported by the South African Research Chairs Initiative of the Department of Science and Technology and the National Research Foundation. HJRVZ is supported in part by the “Quantum Technologies for Sustainable Development” grant from the National Institute for Theoretical and Computational Sciences of South Africa (NITHECS). RdMK and PR would like to thank the Isaac Newton Institute for Mathematical Sciences for support and hospitality during the programme ``Black holes: bridges between number theory and holographic quantum information'' when work on this paper was initiated. This work was also supported by EPSRC Grant Number EP/R014604/1. 

\begin{appendix}

\section{AdS mass operator}\label{App1}

In Section \ref{HMapp} we  have derived formulas for the bulk coordinates $X$, $Z$ and $P^+$ in terms of the conformal field theory coordinates appearing in the trilocal collective field. To obtain the AdS mass operator, given in (\ref{AdSMassOp}) we will need the spin generator $M^{ZX}$, as well as the $B^Z$ operator. In this Appendix we evaluate $M^{ZX}$ and $B^Z$.
 
The bulk higher spin fields are collected into a single field
\bea
\Phi(X^+,X^-,X,Z,\alpha^I)&=&\sum_{s=0}^\infty\alpha_{I_1}\alpha_{I_2}\cdots \alpha_{I_{2s}}{A^{I_1 I_2\cdots I_{2s}}(X^+,X^-,X,Z)\over Z}|0\rangle
\eea
The index $I$ on the oscillators runs over $Z$ and $X$. The trilocal collective field can be decomposed as
\bea
\sigma_3=\sigma_3^0+\eta_3
\eea
where $\sigma_3^0$ is the large $N$ expectation value of $\sigma_3$\footnote{For the free theory this contribution vanishes. With cubic interactions, it will be non-zero.} and $\eta_3$ is a fluctuation about this large $N$ value. It is $\eta_3$ that is mapped to the bulk higher spin gravity field. The holographic mapping between the fields in this case is
\bea
\Phi(X^+,P^+,X,Z,\alpha^I)&=&\mu(p_i^+,x_i)\eta_3(x^+,p_1^+,x_1,p_2^+,x_2,p_3^+,x_3)\label{matchfields}
\eea
where $\mu(p_i^+,x_i)$ is a factor needed to ensure that the conformal generators of the collective field theory and those of the higher spin gravity map into each other under the change of spacetime coordinates given in Section \ref{HMapp}. Using $L_{\rm AdS}$ to denote a bulk generator and $L_{\rm CFT}$ to denote the corresponding collective generator, we have
\bea
L_{\rm AdS}=\mu(p_i^+,x_i) L_{\rm CFT}{1\over\mu(p_i^+,x_i)}\label{LeqL}
\eea
By matching generators we learn that
\bea
\mu(p_i^+,x_i)=\sqrt{p_1^+ p_2^+ p_3^+ \sqrt{p_1^+ p_2^+ (x_1-x_2)^2+p_1^+ p_3^+ (x_1-x_3)^2+p_2^+ p_3^+ (x_2-x_3)^2}}\label{measurefactor}
\eea

To obtain the spin generator, note that by evaluating the conformal generator
\bea
K^X_{\rm AdS}&=&-{1\over 2}(2X^+X^-+X^2+Z^2)\partial^X+X D+M^{XZ}Z+M^{X-}X^+
\eea
at $X^+=0=X$ we obtain
\bea
K^X_{\rm AdS}&=&-{1\over 2}Z^2 \partial^X+M^{XZ}Z
\eea
By using (\ref{LeqL}) to equate this expression for $K^X_{\rm AdS}$ to the generator of the conformal field theory
\bea
K^X_{\rm CFT}&=&-{1\over 2}\sum_{i=1}^3 \left( x_i^2{\partial\over \partial x_i}+x_i D_i\right)
\qquad
D_i\,\,=\,\, -p_i^+{\partial\over\partial p_i^+}+x_i{\partial\over\partial x_i}-{1\over 2}
\eea
we are able to express $M^{XZ}$ in terms of the conformal field theory coordinates. The result is
\bea
M^{XZ}&=&{\sum_{i,j=1}^3 p_j^+(x_i-x_j)^2{\partial\over\partial x_i}-2\sum_{i=1}^3p_i^+v_i{\partial\over\partial p_i^+}\over 2\sqrt{\sum_{i>j=1}^3p_i^+p_j^+(x_i-x_j)^2}}
\eea
In obtaining this result we use the fact that $M^{XZ}$ is translation invariant so it can only depend on coordinate differences and the fact that since $X=0$ we can replace 
\bea
x_i\to {v_i\over p_1^++p_2^++p_3^+}\qquad i=1,2,3
\eea

Next, by evaluating the generator
\bea
K_{\rm AdS}^-&=&-{1\over 2}(2X^+X^-+X^2+Z^2)P^-+X^- D+{1\over\partial^+}X^I\partial^J M^{IJ}\cr\cr
&&-{X\over 2Z\partial^+}[M^{ZX},A]+{1\over\partial^+}B
\eea
we find that
\bea
B&=&(M^{XZ})^2+[M^{XZ},\kappa]^2-{1\over 4}\kappa^2
\eea
where
\bea
\kappa&=&\sqrt{p_1^+p_2^+p_3^+\over p_1^++p_2^++p_3^+}\left({x_2-x_3\over p_1^+}{\partial\over\partial x_1}+{x_3-x_1\over p_2^+}{\partial\over\partial x_2}+{x_1-x_2\over p_3^+}{\partial\over\partial x_3}\right)
\eea 
we are able to express $B$ in terms of the conformal field theory coordinates. Note that $\kappa$, $M^{XZ}$ and $[M^{XZ},\kappa]$ generate an su(2) algebra. Finally, using this expression for $B$ we are able to verify that
\bea
{1\over \mu(p_i^+,x_i)}\Big(Z^2(2P^+P^-+\partial_X^2+\partial_Z^2)-2B-2\Big)\mu(p_i^+,x_i)=C_2
\eea
with $C_2$ the quadratic Casimir defined in defined in (\ref{quadcas}) in terms of the generators of the conformal field theory. With this last identity it is now simple to verify (\ref{CFTeom}).

\section{Generators of the conformal group}\label{App2}

A highly non-trivial confirmation of the holographic mapping proposed in this article is the fact that the generators of the conformal field theory agrees with those of the dual higher spin gravity. This amounts to demonstrating (\ref{LeqL}), after the change of coordinates given in Section \ref{HMapp} and the relation between bulk and boundary fields given in  (\ref{matchfields}) and (\ref{measurefactor}) are used. In this Appendix we list the generators of the conformal field theory and those of the higher spin gravity, that we used to demonstrate (\ref{LeqL}).

\subsection{Conformal Field Theory generators}

The generators we quote below are the generators acting on the trilocal field in the collective field theory. They are obtained using the coproduct
\begin{equation}
\Delta (L)=L\otimes 1\otimes 1+1\otimes L\otimes 1+1\otimes 1\otimes L
\end{equation}
of the standard representation of the generators $L$ of the free scalar field. The only slightly unusual feature is that we have performed a Fourier transform from $x^-$ to $p^+$. The generators are
\begin{eqnarray}
P^+&=&\sum_{i=1}^3 p_i^+\cr\cr
P^x&=&\sum_{i=1}^3{\partial\over \partial x_i}\cr\cr
P^-&=&-\sum_{i=1}^3 {1\over 2p_i^+}{\partial^2\over\partial x_i^2}\cr\cr
J^{+-}&=&x^+ P^- +\sum_{i=1}^3 {\partial\over\partial p_i^+}\, p_i^+\cr\cr
J^{+x}&=&x^+ \sum_{i=1}^3{\partial\over\partial x_i}-\sum_{i=1}^3 x_i p_i^+\cr\cr
J^{-x}&=&\sum_{i=1}^3\left(-{\partial\over\partial p_i^+}{\partial\over\partial x_i}+{x_i\over 2p_i^+}{\partial^2\over\partial x_i^2}\right)\cr\cr
D&=&x^+P^- +\sum_{i=1}^3\left(-{\partial\over\partial p_i^+}p_i^+
+x_i {\partial\over\partial x_i}\right)+{3\over 2}\cr\cr
K^+&=&-{1\over 2}\sum_{i=1}^3\left(-2 x^+ {\partial\over\partial p_i^+}p_i^+
+x_i^2 p_i^+\right)+x^+D\cr\cr
K^-&=& \sum_{i=1}^3\Bigg({3\over 2}{\partial\over\partial p_i^+}
+p_i^+ {\partial^2\over\partial p_i^{+\,\,2}}
-x_i{\partial\over\partial x_i}{\partial\over\partial p_i^+} 
+{x_i^2\over 4p_i^+}{\partial^2\over\partial x_i^2}\Bigg)\cr\cr
K^x&=&-{1\over 2}\sum_{i=1}^3\left(-2 x^+ {\partial\over\partial x_i}{\partial\over\partial p_i^+}+x_i^2 {\partial\over\partial x_i}\right)\cr
&&+\sum_{i=1}^3 x_i \left(-x^+ {1\over 2p_i^+}{\partial^2\over\partial x_i^2}
-{\partial\over\partial p_i^+}p_i^+ + x_i{\partial\over\partial x_i}+{1\over 2}\right)
\end{eqnarray}

\subsection{AdS Generators}

Once again, we rely on results obtained by Metsaev \cite{Metsaev:2003cu}. The formulas below are obtained by Fourier transforming (from $X^-$ to $P^+$) the generators given in Section 2 of \cite{Metsaev:2003cu}. The generators are
\bea
P^X&=&\partial_X\cr\cr
P^+&=&P^+\cr\cr
P^-&=&-{\partial_X^2+\partial_Z^2\over 2P^+}+{1\over 2Z^2P^+}A\cr\cr
D&=&X^+ P^- -P^+\partial_{P^+}+X\partial_X+Z\partial_Z\cr\cr
J^{+-}&=&X^+P^-+P^+\partial_{P^+}+1\cr\cr
J^{+X}&=&X^+\partial_X-XP^+\cr\cr
J^{-X}&=&-\partial_{P^+}\partial_X-XP^-+M^{-X}\cr\cr
K^+&=&-{1\over 2}(X^2+Z^2-2X^+\partial_{P^+})P^++X^+ D\cr\cr
K^X&=&-{1\over 2}(X^2+Z^2-2X^+\partial_{P^+})\partial_X+XD+M^{XZ}Z+M^{X-}X^+\cr\cr
K^-&=&-{1\over 2}(X^2+Z^2-2X^+\partial_{P^+})P^--\partial_{P^+}D+{1\over P^+}(X\partial_Z-Z\partial_X)M^{XZ}\cr\cr
&&-{X\over 2ZP^+}[M^{ZX},A]+{1\over P^+}B
\eea
where $A$ is defined in (\ref{AdSMassOp}) and
\bea
M^{-X}&=&-M^{X-}\,\,=\,\,M^{XZ}{\partial_Z\over P^+}-{1\over 2ZP^+}[M^{ZX},A]
\eea
\bea
B&=&B^Z+M^{ZX}M^{ZX}
\eea
and $B^Z$ is a component of the vector $B^I=(B^X,B^Z)$ which obeys
\bea
\Big([B^I,B^J]+{1\over 2}M^{IK}M^{KL}M^{LJ}-{1\over 2}M^{JK}M^{KL}M^{LI}&&\cr\cr
-(\Delta(\Delta-3)+s(s+1)+M^{XZ}M^{XZ}+2)M^{IJ}\Big)|\phi\rangle&=&0
\eea
The state $|\phi\rangle$ is defined in (\ref{defphi}). In verifying the equality, we replace the combination $\Delta(\Delta-3)+s(s+1)$ by the quadratic Casimir written in terms of the AdS generators. This then gives us a universal set of generators that can be applied to bulk fields in any SO(2,3) representation. It is these universal generators that agree with the generators of the collective field theory.

\end{appendix}


\begin{thebibliography}{}

\bibitem{Maldacena:1997re}
J.~M.~Maldacena, ``The Large N limit of superconformal field theories and supergravity,''
Adv. Theor. Math. Phys. \textbf{2} (1998), 231-252 doi:10.1023/A:1026654312961
[arXiv:hep-th/9711200 [hep-th]].

\bibitem{Gubser:1998bc}
S.~S.~Gubser, I.~R.~Klebanov and A.~M.~Polyakov, ``Gauge theory correlators from noncritical string theory,'' Phys. Lett. B \textbf{428} (1998), 105-114
doi:10.1016/S0370-2693(98)00377-3 [arXiv:hep-th/9802109 [hep-th]].

\bibitem{Witten:1998qj}
E.~Witten, ``Anti-de Sitter space and holography,'' Adv. Theor. Math. Phys. \textbf{2}, 253-291 (1998) doi:10.4310/ATMP.1998.v2.n2.a2 [arXiv:hep-th/9802150 [hep-th]].

\bibitem{Klebanov:2002ja}
I.~R.~Klebanov and A.~M.~Polyakov, ``AdS dual of the critical O(N) vector model,''
Phys. Lett. B \textbf{550} (2002), 213-219 doi:10.1016/S0370-2693(02)02980-5
[arXiv:hep-th/0210114 [hep-th]].

\bibitem{Vasiliev:1990en}
M.~A.~Vasiliev, ``Consistent equation for interacting gauge fields of all spins in (3+1)-dimensions,'' Phys. Lett. B \textbf{243} (1990), 378-382 doi:10.1016/0370-2693(90)91400-6

\bibitem{Das:2003vw}
S.~R.~Das and A.~Jevicki, ``Large N collective fields and holography,'' Phys. Rev. D \textbf{68} (2003), 044011 doi:10.1103/PhysRevD.68.044011 
[arXiv:hep-th/0304093 [hep-th]].

\bibitem{Jevicki:1979mb}
A.~Jevicki and B.~Sakita, ``The Quantum Collective Field Method and Its Application to the Planar Limit,'' Nucl. Phys. B \textbf{165} (1980), 511 doi:10.1016/0550-3213(80)90046-2

\bibitem{Jevicki:1980zg}
A.~Jevicki and B.~Sakita, ``Collective Field Approach to the Large $N$ Limit: Euclidean Field Theories,'' Nucl. Phys. B \textbf{185} (1981), 89-100 doi:10.1016/0550-3213(81)90365-5

\bibitem{deMelloKoch:2010wdf}
R.~de Mello Koch, A.~Jevicki, K.~Jin and J.~P.~Rodrigues, ``$AdS_4/CFT_3$ Construction from Collective Fields,'' Phys. Rev. D \textbf{83} (2011), 025006 doi:10.1103/PhysRevD.83.025006 [arXiv:1008.0633 [hep-th]].

\bibitem{Metsaev:1999ui}
R.~R.~Metsaev, ``Light cone form of field dynamics in Anti-de Sitter space-time and AdS / CFT correspondence,'' Nucl. Phys. B \textbf{563} (1999), 295-348 doi:10.1016/S0550-3213(99)00554-4 [arXiv:hep-th/9906217 [hep-th]].

\bibitem{deMelloKoch:2023ngh}
R.~de Mello Koch, ``Microscopic entanglement wedges,'' JHEP \textbf{08} (2023), 056
doi:10.1007/JHEP08(2023)056 [arXiv:2307.05032 [hep-th]].

\bibitem{Mintun:2014gua}
E.~Mintun and J.~Polchinski, ``Higher Spin Holography, RG, and the Light Cone,''
[arXiv:1411.3151 [hep-th]].

\bibitem{deMelloKoch:2021cni}
R.~de Mello Koch, E.~Gandote, N.~H.~Tahiridimbisoa and H.~J.~R.~Van Zyl,
``Quantum error correction and holographic information from bilocal holography,''
JHEP \textbf{11} (2021), 192 doi:10.1007/JHEP11(2021)192 [arXiv:2106.00349 [hep-th]].

\bibitem{Dong:2016eik}
X.~Dong, D.~Harlow and A.~C.~Wall, ``Reconstruction of Bulk Operators within the Entanglement Wedge in Gauge-Gravity Duality,'' Phys. Rev. Lett. \textbf{117} (2016) no.2, 021601 doi:10.1103/PhysRevLett.117.021601 [arXiv:1601.05416 [hep-th]].

\bibitem{deMelloKoch:2022sul}
R.~de Mello Koch and G.~Kemp, ``Holography of information in AdS/CFT,''
JHEP \textbf{12} (2022), 095 doi:10.1007/JHEP12(2022)095 [arXiv:2210.11066 [hep-th]].

\bibitem{Laddha:2020kvp}
A.~Laddha, S.~G.~Prabhu, S.~Raju and P.~Shrivastava, ``The Holographic Nature of Null Infinity,'' SciPost Phys. \textbf{10} (2021) no.2, 041 doi:10.21468/SciPostPhys.10.2.041
[arXiv:2002.02448 [hep-th]].

\bibitem{deMelloKoch:2024juz}
R.~de Mello Koch, G.~Kemp and H.~J.~R.~Van Zyl, ``Bilocal holography and locality in the bulk,'' [arXiv:2403.07606 [hep-th]].

\bibitem{Jevicki:2011ss}
A.~Jevicki, K.~Jin and Q.~Ye, ``Collective Dipole Model of AdS/CFT and Higher Spin Gravity,'' J. Phys. A \textbf{44} (2011), 465402 doi:10.1088/1751-8113/44/46/465402 [arXiv:1106.3983 [hep-th]].

\bibitem{Jevicki:2011aa}
A.~Jevicki, K.~Jin and Q.~Ye, ``Bi-local Model of AdS/CFT and Higher Spin Gravity,''
[arXiv:1112.2656 [hep-th]].

\bibitem{deMelloKoch:2017caf}
R.~de Mello Koch, P.~Rabambi, R.~Rabe and S.~Ramgoolam, ``Free quantum fields in 4D and Calabi-Yau spaces,'' Phys. Rev. Lett. \textbf{119} (2017) no.16, 161602
doi:10.1103/PhysRevLett.119.161602 [arXiv:1705.04039 [hep-th]].

\bibitem{deMelloKoch:2017dgi}
R.~de Mello Koch, P.~Rabambi, R.~Rabe and S.~Ramgoolam, ``Counting and construction of holomorphic primary fields in free CFT4 from rings of functions on Calabi-Yau orbifolds,''
JHEP \textbf{08} (2017), 077 doi:10.1007/JHEP08(2017)077 [arXiv:1705.06702 [hep-th]].

\bibitem{DeMelloKoch:2018hyq}
R.~de Mello Koch, P.~Rambambi and H.~J.~R.~Van Zyl, ``From Spinning Primaries to Permutation Orbifolds,'' JHEP \textbf{04} (2018), 104 doi:10.1007/JHEP04(2018)104
[arXiv:1801.10313 [hep-th]].

\bibitem{deMelloKoch:2018klm}
R.~de Mello Koch and S.~Ramgoolam, ``Free field primaries in general dimensions: Counting and construction with rings and modules,'' JHEP \textbf{08} (2018), 088
doi:10.1007/JHEP08(2018)088 [arXiv:1806.01085 [hep-th]].

\bibitem{Bae:2016hfy}
J.~B.~Bae, E.~Joung and S.~Lal, ``On the Holography of Free Yang-Mills,''
JHEP \textbf{10} (2016), 074 doi:10.1007/JHEP10(2016)074 [arXiv:1607.07651 [hep-th]].

\bibitem{Dolan:2005wy}
F.~A.~Dolan, ``Character formulae and partition functions in higher dimensional conformal field theory,'' J. Math. Phys. \textbf{47} (2006), 062303 doi:10.1063/1.2196241
[arXiv:hep-th/0508031 [hep-th]].

\bibitem{Flato:1978qz}
M.~Flato and C.~Fronsdal, ``One Massless Particle Equals Two Dirac Singletons: Elementary Particles in a Curved Space. 6.,'' Lett. Math. Phys. \textbf{2} (1978), 421-426
doi:10.1007/BF00400170

\bibitem{Basile:2014wua}
T.~Basile, X.~Bekaert and N.~Boulanger, ``Flato-Fronsdal theorem for higher-order singletons,'' JHEP \textbf{11} (2014), 131 doi:10.1007/JHEP11(2014)131 [arXiv:1410.7668 [hep-th]].

\bibitem{Giombi:2016hkj}
S.~Giombi and V.~Kirilin, ``Anomalous dimensions in CFT with weakly broken higher spin symmetry,'' JHEP \textbf{11} (2016), 068 doi:10.1007/JHEP11(2016)068 [arXiv:1601.01310 [hep-th]].

\bibitem{deMelloKoch:2022dpj}
R.~de Mello Koch and S.~Ramgoolam, ``$\mathcal{N}=4$ SYM, (super)-polynomial rings and emergent quantum mechanical symmetries,'' [arXiv:2211.04271 [hep-th]].

\bibitem{Newton:2008au}
T.~H.~Newton and M.~Spradlin, ``Quite a Character: The Spectrum of Yang-Mills on S3,''
Phys. Lett. B \textbf{672} (2009), 382-385 doi:10.1016/j.physletb.2009.01.044
[arXiv:0812.4693 [hep-th]].

\bibitem{Metsaev:2003cu}
R.~R.~Metsaev, ``Massive totally symmetric fields in AdS(d),'' Phys. Lett. B \textbf{590} (2004), 95-104 doi:10.1016/j.physletb.2004.03.057 [arXiv:hep-th/0312297 [hep-th]].

\bibitem{Metsaev:2005ws}
R.~R.~Metsaev, ``Light-cone formulation of conformal field theory adapted to AdS/CFT correspondence,'' Phys. Lett. B \textbf{636} (2006), 227-233 doi:10.1016/j.physletb.2006.03.052 [arXiv:hep-th/0512330 [hep-th]].

\bibitem{Metsaev:2013kaa}
R.~R.~Metsaev, ``Light-cone gauge approach to arbitrary spin fields, currents, and shadows,'' J. Phys. A \textbf{47} (2014), 375401 doi:10.1088/1751-8113/47/37/375401
[arXiv:1312.5679 [hep-th]].

\bibitem{Metsaev:2015rda}
R.~R.~Metsaev, ``Light-cone AdS/CFT-adapted approach to AdS fields/currents, shadows, and conformal fields,'' JHEP \textbf{10} (2015), 110 doi:10.1007/JHEP10(2015)110
[arXiv:1507.06584 [hep-th]].

\bibitem{Metsaev:2022ndg}
R.~R.~Metsaev, ``Light-cone gauge massive and partially-massless fields in AdS(4),''
Phys. Lett. B \textbf{839} (2023), 137790 doi:10.1016/j.physletb.2023.137790
[arXiv:2212.14728 [hep-th]].

\bibitem{Chowdhury:2020hse}
C.~Chowdhury, O.~Papadoulaki and S.~Raju, ``A physical protocol for observers near the boundary to obtain bulk information in quantum gravity,'' SciPost Phys. \textbf{10} (2021) no.5, 106 doi:10.21468/SciPostPhys.10.5.106 [arXiv:2008.01740 [hep-th]].

\bibitem{Raju:2020smc}
S.~Raju, ``Lessons from the information paradox,'' Phys. Rept. \textbf{943} (2022), 1-80
doi:10.1016/j.physrep.2021.10.001 [arXiv:2012.05770 [hep-th]].

\bibitem{Raju:2021lwh}
S.~Raju,``Failure of the split property in gravity and the information paradox,''
Class. Quant. Grav. \textbf{39} (2022) no.6, 064002 doi:10.1088/1361-6382/ac482b
[arXiv:2110.05470 [hep-th]].

\bibitem{SuvratYouTube}
For a beautiful set of lectures, incredibly helpful when learning this material, go to: https://www.youtube.com/channel/UCJ-YA8uOwUlACfn49iD7TvA

\bibitem{Marolf:2008mf}
D.~Marolf, ``Unitarity and Holography in Gravitational Physics,'' Phys. Rev. D \textbf{79} (2009), 044010 doi:10.1103/PhysRevD.79.044010 [arXiv:0808.2842 [gr-qc]].

\bibitem{Jacobson:2012ubm}
T.~Jacobson, ``Boundary unitarity and the black hole information paradox,'' Int. J. Mod. Phys. D \textbf{22} (2013), 1342002 doi:10.1142/S0218271813420029 [arXiv:1212.6944 [hep-th]].

\bibitem{Papadodimas:2012aq}
K.~Papadodimas and S.~Raju, ``An Infalling Observer in AdS/CFT,'' JHEP \textbf{10} (2013), 212 doi:10.1007/JHEP10(2013)212 [arXiv:1211.6767 [hep-th]].

\bibitem{Banerjee:2016mhh}
S.~Banerjee, J.~W.~Bryan, K.~Papadodimas and S.~Raju, ``A toy model of black hole complementarity,'' JHEP \textbf{05} (2016), 004 doi:10.1007/JHEP05(2016)004 [arXiv:1603.02812 [hep-th]].

\bibitem{Raju:2018zpn}
S.~Raju, ``A Toy Model of the Information Paradox in Empty Space,'' SciPost Phys. \textbf{6} (2019) no.6, 073 doi:10.21468/SciPostPhys.6.6.073 [arXiv:1809.10154 [hep-th]].

\bibitem{Raju:2019qjq}
S.~Raju, ``Is Holography Implicit in Canonical Gravity?,'' Int. J. Mod. Phys. D \textbf{28} (2019) no.14, 1944011 doi:10.1142/S0218271819440115 [arXiv:1903.11073 [gr-qc]].

\bibitem{Jacobson:2019gnm}
T.~Jacobson and P.~Nguyen, ``Diffeomorphism invariance and the black hole information paradox,'' Phys. Rev. D \textbf{100} (2019) no.4, 046002 doi:10.1103/PhysRevD.100.046002
[arXiv:1904.04434 [gr-qc]].


\end{thebibliography}
\end{document}